\documentclass[aps,prl,twocolumn,showpacs]{revtex4}
\usepackage{graphicx}
\usepackage{natbib}

\begin{document}

\title{Spatially heterogeneous dynamics and dynamic facilitation in a model of viscous silica}

\author{Michael Vogel}
\email{mivogel@umich.edu}
\author{Sharon C. Glotzer}
\affiliation{Departments of Chemical Engineering and Materials
Science and Engineering, University of Michigan, 2300 Hayward, Ann
Arbor, MI, 48109, USA}

\date{\today}

\begin{abstract}
Performing molecular dynamics simulations, we find that the
structural relaxation dynamics of viscous silica, the prototype of a
strong glass former, are spatially heterogeneous and cannot be
understood as a statistical bond breaking process. Further, we show
that high particle mobility predominantly propagates continuously
through the melt, supporting the concept of dynamic facilitation
emphasized in recent theoretical work.
\end{abstract}

\maketitle

One of the most challenging problems of condensed matter physics is
still the understanding of the tremendous slowing down of molecular
dynamics in supercooled liquids approaching the glass transition
temperature $T_g$, which is not accompanied by a substantial change
of the structure. In the past, several theories have been invoked to
rationalize this behavior. The mode coupling theory~\cite{MCT}
describes many experimental findings at high temperatures, but it
predicts a power-law divergence of relaxation times at a critical
temperature $T_{MCT}$ significantly above $T_g$. The Adam-Gibbs
approach~\cite{AG} sees the glass transition phenomenon as resulting
from an increase of cooperativity of molecular dynamics when the
temperature is decreased. However, the underlying microscopic picture 
is not well established.

Very recently, Garrahan and Chandler (GC)~\cite{GC-PRL,GC-PN}
introduced a microscopic model of supercooled liquids, which is based
on three central ideas: (i) Particle mobility is sparse and dynamics
are spatially heterogeneous at times intermediate between ballistic
and diffusive motion. (ii) Particle mobility is the result of dynamic
facilitation, i.e., mobile particles assist their neighbors to become
mobile. (iii) Mobility propagation carries a direction, the
persistence length of which is larger for fragile than for strong
glass formers~\cite{AA}. For non-strong liquids, the existence of
spatially heterogeneous dynamics (SHD) is well established by
experiment~\cite{AH-4D,HS-RE,ME-RE} and
simulation~\cite{DH-AH,DH-PH,DH-CD1,DH-CB,DH-CD2,DH-YG-PO,DH-YG-DZ,DH-MA,DH-H2O}.
Further, it was observed~\cite{DH-CD1,DH-CD2,DH-YG-DZ,DH-MA} that
groups of particles follow one another along string-like paths. GC
proposed~\cite{GC-PN} that dynamic facilitation and persistence of
particle-flow direction manifest themselves in this string-like
motion. For strong liquids, like silica, experimental studies
are rare due to the high $T_g$ of these network-forming
materials. However, it is usually argued that the structural
relaxation can be understood as simple activated bond breaking
process.

Here, we perform molecular dynamics (MD) simulations to study the
relaxation dynamics of viscous silica, the prototype of a strong
glass former~\cite{AA}. In particular, we critically test the ideas
of the GC model. For our simulations, we apply the BKS
potential~\cite{BKS}, which is commonly used to reproduce structural
and dynamical properties of viscous
silica~\cite{JB,KV,JH-BP1,JH-ALL,JH-FS,SV}. We show that the
relaxation dynamics in BKS silica are spatially heterogeneous,
indicating a complex nature of this dynamical process. Further, we
demonstrate quantitatively that dynamic facilitation is not only
relevant in spin models~\cite{FA}, but also in a viscous equilibrium
liquid. Finally, we show that string-like motion is less important in 
BKS silica than in non-strong liquids, suggesting a
reduced persistence of particle-flow direction, in agreement with the
GC model. To arrive at these results, we follow Horbach and
Kob~\cite{JH-ALL,JH-FS} and perform simulations in the NVE ensemble
with $\rho\!=\!2.37\,\mathrm{g/cm^3}$ and $N\!=\!8016$, which is
sufficiently large so as to avoid finite size effects at the studied
temperatures. Further simulation details can be found in
Refs.~\cite{JH-ALL,JH-FS,MV-PRB}.

\begin{figure}
\includegraphics[angle=0,width=7.8cm]{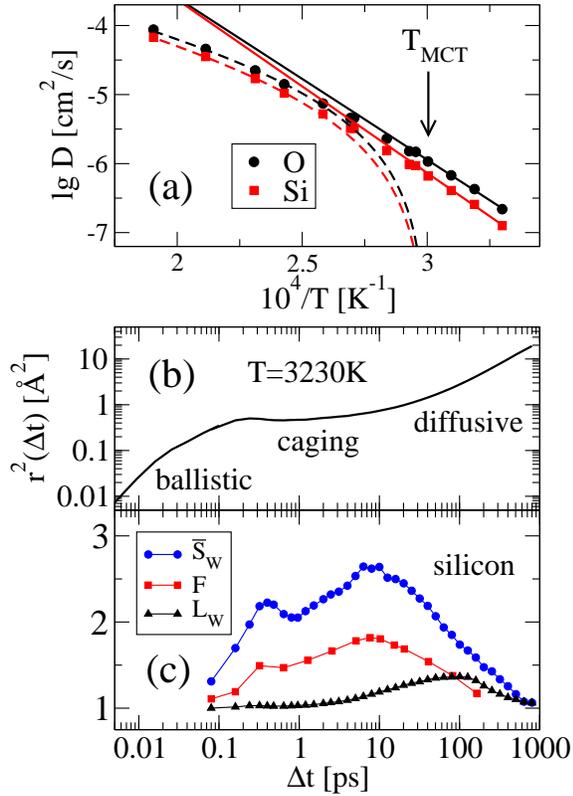}
\caption{(a) Diffusion coefficients $D(1/T)$ for the oxygen and
silicon atoms. Dashed lines: Power-laws
$D\!\propto\!(T\!-\!3330\mathrm{\,K})^{\gamma}$ with $\gamma\!=\!2.0$
and $\gamma\!=\!2.1$ for the oxygen and silicon atoms, respectively.
Solid lines: Arrhenius laws with activation energies
$E_a\!=\!4.7\mathrm{\,eV}$ for oxygen and $E_a\!=\!5.0\mathrm{\,eV}$
for silicon. (b) Mean square displacement, $r^2(\Delta t)$, for the
silicon atoms at $T\!=\!3230\mathrm{\,K}$. (c) Normalized mean
cluster size $\overline{S}_W(\Delta t)$, mean string length
$L_W(\Delta t)$ and the dynamic facilitation effect $F(\Delta t)$,
cf.\ Eq.~\ref{F}, for the silicon atoms at
$T\!=\!3230\mathrm{\,K}$.}\label{fig1}
\end{figure}

Consistent with experimental results~\cite{ER}, the temperature
dependence of the transport coefficients in BKS silica changes from a
non-Arrhenius behavior at high $T$ to an Arrhenius behavior at low
$T$, where the crossover occurs in the vicinity of
$T_{MCT}$~\cite{JH-ALL,JH-FS,SV}. This result is confirmed by our
analysis, see Fig.~\ref{fig1}(a). Using the literature value
$T_{MCT}\!=\!3330\mathrm{\,K}$~\cite{JH-ALL,JH-FS}, we find that the
diffusion coefficients $D$ for both oxygen and silicon vary according
to a power-law~\cite{MCT}, $D\!\propto\!(T\!-\!T_{MCT})^{\gamma}$, at
high $T$, whereas Arrhenius laws are observed at
$T\!\lesssim\!T_{MCT}$. The activation energies
$E_a\!=\!4.7\mathrm{eV}$ and $E_a\!=\!5.0\mathrm{eV}$ for the oxygen
and silicon atoms, respectively, are consistent with previous results
from simulations~\cite{JH-ALL,JH-FS} and experiments~\cite{EAO,EAS}.
Following the time evolution of the mean square displacement,
$r^2(\Delta t)$, in Fig.~\ref{fig1}(b), three distinct time regimes
can be distinguished at sufficiently low $T$~\cite{JH-ALL}. Ballistic
and diffusive motion at short and long times, respectively, are
separated by a pronounced caging regime~\cite{MCT}, where the
particles are temporarily trapped in cages formed by their neighbors.

To demonstrate that dynamics in silica are spatially heterogenous, we
show that highly mobile particles form clusters larger than predicted
from random statistics. Following previous
work~\cite{DH-CD2,DH-YG-PO,DH-H2O,DH-YG-DZ}, we characterize the
particle mobility in a time interval $\Delta t$ by the scalar
displacement and select the 5\% most mobile particles for further
analysis. Then, we define a cluster as a group of the most mobile
particles that reside in the first neighbor shell of each other. From
the probability distribution $P_S(n;\Delta t)$ of finding a cluster
of size $n$ within a time interval $\Delta t$ we calculate the
weight-averaged mean cluster size
\begin{equation}\label{WA} S_{W}(\Delta
t)=\frac{\sum_n\,n^2\,P_S(n;\Delta t)}{\sum_n\,n\,P_S(n;\Delta t)}.
\end{equation}
This quantity measures the average size of a cluster to which a
mobile particle belongs. To eliminate the random contribution, we
discuss the normalized size $\overline{S}_W(\Delta t)\!=\!S_W(\Delta
t)/S_W^*$, where $S_W^*$ is the value that results when 5\% of the
particles are selected irrespective of their mobility. Our
conclusions are unchanged, when the fraction of mobile particles is
varied over a meaningful range~\cite{MV-PRB}.

\begin{figure}
\includegraphics[angle=270,width=8cm]{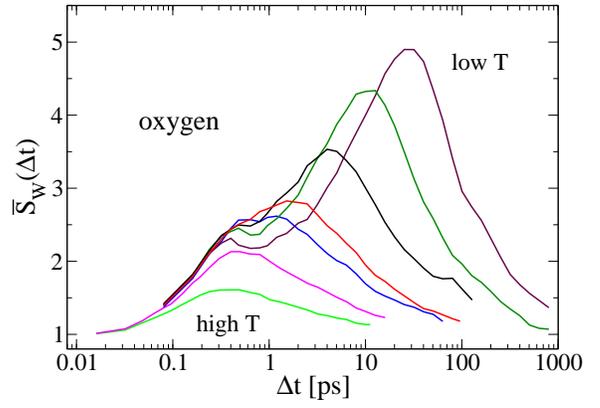}
\caption{Normalized mean cluster size, $\overline{S}_W(\Delta t)$,
for the oxygen atoms at various temperatures ($5250\,\mathrm{K}$,
$4330\,\mathrm{K}$, $3870\,\mathrm{K}$, $3690\,\mathrm{K}$,
$3390\,\mathrm{K}$, $3230\,\mathrm{K}$, $3030\,\mathrm{K}$).}\label{fig2}
\end{figure}

Figure~\ref{fig1}(c) shows $\overline{S}_W(\Delta t)$ for the silicon
atoms at $T\!=\!3230\mathrm{\,K}$. We see that the mean cluster size
exhibits two maxima, indicating the existence of two spatially 
heterogeneous dynamical processes. The secondary and primary
maxima are located at the crossovers from the ballistic to the caging
regime and from the caging to the diffusive regime, respectively. In
Fig.~\ref{fig2}, we depict the temperature dependence of
$\overline{S}_W(\Delta t)$ for the oxygen atoms. The behavior of the
primary maximum is analogous to that for simple liquids and polymer
melts~\cite{DH-YG-DZ,DH-YG-PO}. Both the peak time $t_S$ and the peak
height $\overline{S}_W(t_S)$ increase upon cooling. Here, the growth
of the clusters continues in the temperature range where $D$ follows
an Arrhenius law, cf.\ Fig.~\ref{fig1}(a). For $T\!\approx\!T_{MCT}$,
the mean cluster sizes $\overline{S}_W(t_S)$ for the oxygen and silicon 
are located at the lower end of the spectrum of values reported for
non-strong liquids~\cite{DH-CD2,DH-YG-DZ,DH-YG-PO,DH-H2O}. Hence, at 
times when the particles escape from their cages, BKS silica exhibits 
SHD, which is at least qualitatively comparable to that of non-strong liquids. 
The position of the secondary maximum is temperature independent so 
that both peaks merge for $T\!\gtrsim4000\mathrm{\,K}$. In studies of
water~\cite{DH-H2O}, an analogous secondary peak was observed and
ascribed to ``strong correlations in the vibrational motion of
first-neighbor molecules''. For BKS silica, a Boson peak was found at
the crossover from the ballistic to the caging regime~\cite{JH-BP1}.
Therefore, we speculate that this phenomenon leads to the secondary
peak and, hence, can be characterized as localized, cooperative
vibrational motion.

String-like motion was found to be an important channel of relaxation
in fragile liquids~\cite{DH-CD1,DH-CD2,DH-MA,DH-YG-DZ}. To study this
type of motion for silica, we follow Donati \emph{et al.\
}\cite{DH-CD1} and construct strings by connecting any two particles
$i$ and $j$ of the same atomic species if
\begin{displaymath}
\mathrm{min}[\,|\vec{r}_{i}(t_0)\!-\!\vec{r}_{j}(t_0\!+\!\Delta
t)|,|\vec{r}_{i}(t_0\!+\!\Delta t)\!-\!\vec{r}_{j}(t_0)|\,]
\!<\!\delta.
\end{displaymath}
is valid for the particle positions at two different times. To ensure
that one particle has moved and another particle has occupied its
position, $\delta$ is set to about 55\% of the respective
interparticle distance. In Fig.~\ref{fig1}(c), we include the
weight-averaged mean string length $L_W(\Delta t)$, calculated for
the silicon atoms at $T\!=\!3230\mathrm{\,K}$ in analogy with
Eq.~\ref{WA}. From the small values $L_W(\Delta t)\!\approx\!1$, it is
clear that string-like motion is of little relevance for the
relaxation of the silicon atoms. Though the strings are somewhat
longer for the oxygen atoms, e.g. a maximum value $L_W(t_L)\!=\!1.5$ 
is found at $T\!=\!3230\mathrm{\,K}$, they are still much smaller 
than in non-strong liquids, where $L_W(t_L)\!>\!3$ for 
$T\!\approx\!T_{MCT}$~\cite{DH-MA,DH-YG-DZ}. Thus, we conclude 
that string-like motion in BKS silica is suppressed due to
the presence of covalent bonds at the studied $T$.

\begin{figure}
\includegraphics[angle=270,width=8cm]{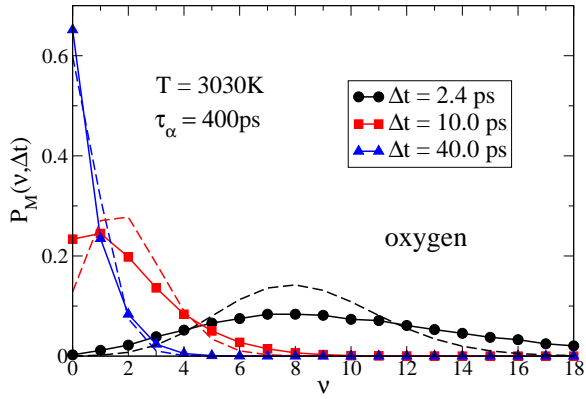}
\caption{Probability distributions $P_M(\nu,\Delta t)$ of finding an
oxygen atom $\nu$ times mobile when dividing $\tau_{\alpha}$ into
time intervals $\Delta t$ and identifying the mobile particles in
each interval. The dashed lines are the corresponding expectations,
$P_M^*(\nu,\Delta t)$, for a random selection of mobile particles in
each interval. At the studied temperature $T\!=\!3030\mathrm{\, K}$,
we find $t_S\approx\!10\mathrm{\,ps}$.}\label{fig3}
\end{figure}

Having established the existence of dynamic heterogeneities, we now
measure their lifetime. For this purpose, we identify the most mobile
particles in back-to-back time intervals, $\Delta t_{12}\!=\!\Delta
t$ and $\Delta t_{23}\!=\!\Delta t$, and calculate the probability
$P(\Delta t)$ of finding a mobile particle in $\Delta t_{12}$ still
mobile in $\Delta t_{23}$. For both atomic species at
$T\!=\!3030\mathrm{\,K}$, we find a nearly constant value $P(\Delta
t)\!\approx\!0.15$ for $\Delta t$ in the caging regime, followed by a
decrease towards the long-time limit $P(\Delta t
\!\rightarrow\!\infty)\!=\!0.05$. These small values indicate that
the fluctuations of the dynamical state are rapid. For the dynamic
heterogeneities in non-strong liquids, such short lifetimes were
observed both in simulation studies near
$T_{MCT}$~\cite{DH-AH,DH-CD2} and in experimental work near, but not
too close to $T_g$~\cite{HS-RE,ME-RE,HWS}.

Next, we investigate the number of times a particular particle is
mobile on the time scale of the $\alpha$-relaxation time
$\tau_{\alpha}$. For this analysis, we divide $\tau_{\alpha}$ into
several time intervals $\Delta t$, identify the most mobile particles
in each interval, and determine the probability distribution
$P_M(\nu,\Delta t)$ of finding a particle mobile in $\nu$ not
necessarily consecutive intervals. Figure~\ref{fig3} shows this
distribution for the oxygen atoms at $T\!=\!3030\mathrm{\,K}$. For
comparison, we include the distribution $P_M^*(\nu,\Delta t)$,
resulting for a random selection of 5\% of the particles in the
intervals $\Delta t$. For $\Delta t\!\geq\!t_S$, we find
$P_M(\nu)\!\approx\!P_M^*(\nu)$ and, hence, the subsets of mobile
particles in different intervals are statistically independent. Thus,
high particle mobility is not limited to certain regions of the
sample, but all particles can belong to a cluster of mobile particles
on the time scale of $\tau_{\alpha}$. Some deviations between
$P_M(\nu)$ and $P_M^*(\nu)$ for $\Delta t\!<\!t_S$ suggest that the
relaxation of a particular region of the sample is not complete on
this short timescale, consistent with the continuing growth of the
clusters.

\begin{figure}
\includegraphics[angle=270,width=8cm]{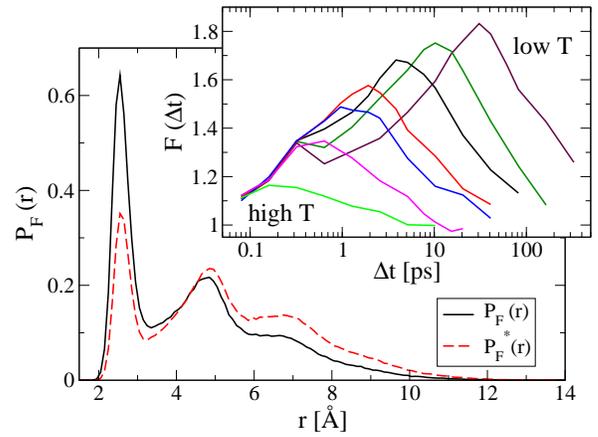}
\caption{Probability distribution $P_F(r,\Delta t)$ of finding a
smallest distance $r$ between an oxygen atom, which becomes mobile in
$\Delta t_{23}\!=\!\Delta t$, and any oxygen atom, which was mobile
in $\Delta t_{12}\!=\!\Delta t$ ($T\!=\!3030\mathrm{\, K})$.
$P_F^*(r,\Delta t)$ results when the mobile particles in $\Delta
t_{23}$ are randomly selected from the non-mobile particles in
$\Delta t_{12}$. The distance is calculated based on the
configuration at the time $t_2$, separating both time intervals. The
inset shows the temperature dependence of $F(\Delta t)$ quantifying
the strength of dynamic facilitation for the oxygen atoms, cf.\
Eq.~\ref{F}. The temperatures are $5250\,\mathrm{K}$,
$4330\,\mathrm{K}$, $3870\,\mathrm{K}$, $3690\,\mathrm{K}$,
$3390\,\mathrm{K}$, $3230\,\mathrm{K}$ and $3030\,\mathrm{K}$.}
\label{fig4}
\end{figure}

Finally, we study whether mobility propagates continuously, as
proposed in the GC model, or whether it develops at random positions,
as suggested by Stillinger and Hodgdon~\cite{FHS} in the model of
flickering fluidized domains. To answer this question, we analyze the
relative positions of particles that are mobile in back-to-back time
intervals $\Delta t_{12}\!=\!\Delta t$ and $\Delta t_{23}\!=\!\Delta
t$, respectively. Specifically, we calculate the probability
distribution $P_F(r,\Delta t)$ of finding a smallest distance $r$
between a particle that is mobile in $\Delta t_{23}$, but not in
$\Delta t_{12}$, and any of the mobile particles in $\Delta t_{12}$.
In Fig.~\ref{fig4}, this distribution is shown for the oxygen atoms
at $T\!=\!3030\mathrm{\,K}$, where $\Delta t\!\approx\!t_S$. For
further analysis, we compare these data with the distribution
$P_F^*(r,\Delta t)$, which results when the particles used for
analysis are randomly selected from the non-mobile oxygen atoms in
$\Delta t_{12}$. We see that the probability that a neighbor of a
mobile oxygen atom becomes mobile in the subsequent time interval is
significantly enhanced. Similar results are observed for the silicon
atoms and, hence, we can conclude that mobile particles in BKS silica
assist their neighbors to become mobile, demonstrating that dynamic
facilitation occurs.

The strength of this effect can be quantified based on the quantity
\begin{equation}\label{F}
F(\Delta t)=\frac{\int_0^{r_{min}}P_F(r,\Delta
t)\,dr}{\int_0^{r_{min}}P_F^*(r,\Delta t)\,dr},
\end{equation}
where $r_{min}$ is the first minimum of the pair correlation
function. This quantity measures by how much the probability that a
neighbor of a mobile particle becomes mobile is enhanced, as compared
to the value for a random development of particle mobility.
Figure~\ref{fig1}(c) displays $F(\Delta t)$ for the silicon atoms at
$T\!=\!3230\mathrm{\,K}$. We see that the effect of dynamic
facilitation is strongest at a time $t_F\!\approx\!t_S$, i.e., when
the clusters of mobile particles are largest. Further,
$F(t_F)\!\approx\!2$ indicates that dynamic facilitation plays a
significant role. The inset of Figure~\ref{fig4} depicts the
temperature dependence of $F(\Delta t)$ for the oxygen atoms. It is clearly seen
that the relevance of dynamic facilitation strongly increases when
$T$ is decreased, suggesting that this effect becomes very important
near $T_g$.

We have shown that the most mobile particles in BKS silica form
clusters that are biggest at the crossover from the caging regime to
the diffusive regime, indicating that this strong glass former
exhibits SHD. Since the most mobile particles show displacements
comparable to the interparticle distance at these
times~\cite{MV-PRB}, we conclude that the structural relaxation of
the studied model liquid is not a result of a simple bond breaking
process occurring at random positions. Further, we have found that
the time and temperature dependence of the mean cluster size are
similar to the behavior for simple
liquids~\cite{DH-CD2,DH-YG-PO,DH-YG-DZ}, implying that strong and
fragile glass formes exhibit no principal, but at most quantitative
differences in the nature of SHD on intermediate timescales. However,
cooperative string-like motion is less relevant in BKS silica than in
non-network forming liquids~\cite{DH-CD1,DH-YG-DZ,DH-MA}, suggesting
that this dynamical pattern is suppressed by the presence of covalent
bonds. Additionally, we have shown that the dynamic heterogeneities
in BKS silica are short-lived and spread throughout the sample on the
time scale of $\tau_{\alpha}$. In doing so, the probability that a
particle becomes mobile is enhanced in the vicinity of another mobile
particle. This tendency for continuous propagation of mobility
demonstrates the relevance of dynamic facilitation. The effect of
dynamic facilitation is strongest at times when the mean cluster size
is maximum and it increases upon cooling, suggesting that this effect
is of particular importance when $T_g$ is approached.

In conclusion, we have shown for a strong glass former that (i)
dynamics at intermediate timescales are spatially heterogeneous, (ii)
dynamic facilitation is important and (iii) the persistence length of
particle-flow direction as measured by the relevance of string-like
motion is smaller than in fragile liquids. These points are central
ideas of the GC model~\cite{GC-PRL,GC-PN} and, hence, our findings
support this new approach to the glass transition phenomenon. In
addition, our results provide insights into the relaxation dynamics
of one of the most important glass formers, which are difficult to
obtain from experimental studies.

\begin{acknowledgments}
M.\ V.\ acknowledges funding of the DFG through the Emmy
Noether-Programm.
\end{acknowledgments}

\end{document}